\documentclass[conference,final]{IEEEtran}
\IEEEoverridecommandlockouts

\usepackage[utf8]{inputenc}
\usepackage[T1]{fontenc}
\usepackage{microtype,newtxtext,newtxmath} 
\usepackage{mathtools,gensymb,eurosym,url,soul} 
\usepackage{subcaption}
\usepackage{longtable,booktabs,array}
\usepackage[backend=biber, sorting=none, style=ieee, maxnames=5]{biblatex}
\usepackage[switch]{lineno}
\usepackage{acronym}
\usepackage{graphicx}
\usepackage{xcolor}
\usepackage{xr} 
\usepackage{textcomp}
\usepackage{hyperref}
\hypersetup{colorlinks=true, breaklinks=true, 
  urlcolor=blue, linkcolor=blue,anchorcolor=blue,citecolor=blue,
  hypertexnames=true, final=true,
  pdfpagemode = UseNone, 
  pdfauthor = {},
  pdftitle = {},
  pdfsubject = {},
  pdfkeywords = {}
}

\usepackage{amsmath}
\usepackage{array}
\usepackage{footnote}
\usepackage{amsmath}

\graphicspath{{figures/} {img/} {./}}
\DeclareGraphicsExtensions{.pdf,.png,.jpg,.mps}

\IEEEdisplaynontitleabstractindextext
\begin{document}
\title{An Efficient Multicast Addressing Encoding Scheme for Multi-Core Neuromorphic Processors\\
{\footnotesize \textsuperscript{}}
\thanks{}
}

\acrodef{ADC}[ADC]{Analog-to-Digital Converter}
\acrodef{ADEXP}[AdExp-IF]{Adaptive Exponential Integrate-and-Fire}
\acrodef{ADM}[ADM]{Asynchronous Delta Modulator}
\acrodef{AE}[AE]{Address-Event}
\acrodef{AER}[AER]{Address-Event Representation}
\acrodef{AEX}[AEX]{AER EXtension board}
\acrodef{AFE}[AFE]{Analog Front-End}
\acrodef{AFM}[AFM]{Atomic Force Microscope}
\acrodef{AGC}[AGC]{Automatic Gain Control}
\acrodef{AI}[AI]{Artificial Intelligence}
\acrodef{AMDA}[AMDA]{AER Motherboard with D/A converters}
\acrodef{AMPA}[AMPA]{$\alpha$-Amino-3-hydroxy-5-methyl-4-isoxazolepropionic Acid}
\acrodef{ANN}[ANN]{Artificial Neural Network}
\acrodef{API}[API]{Application Programming Interface}
\acrodef{APMOM}[APMOM]{Alternate Polarity Metal On Metal}
\acrodef{ARM}[ARM]{Advanced RISC Machine}
\acrodef{ASIC}[ASIC]{Application Specific Integrated Circuit}
\acrodef{BCM}[BMC]{Bienenstock-Cooper-Munro}
\acrodef{BD}[BD]{Bundled Data}
\acrodef{BEOL}[BEOL]{Back-end of Line}
\acrodef{BG}[BG]{Bias Generator}
\acrodef{BMI}[BMI]{Brain-Machince Interface}
\acrodef{BTB}[BTB]{Band-to-Band tunnelling}
\acrodef{CA}[CA]{Cortical Automaton}
\acrodef{CAD}[CAD]{Computer Aided Design}
\acrodef{CAM}[CAM]{Content Addressable Memory}
\acrodef{CAVIAR}[CAVIAR]{Convolution AER Vision Architecture for Real-Time}
\acrodef{CCN}[CCN]{Cooperative and Competitive Network}
\acrodef{CDR}[CDR]{Clock-Data Recovery}
\acrodef{CFC}[CFC]{Current to Frequency Converter}
\acrodef{CHP}[CHP]{Communicating Hardware Processes}
\acrodef{CMIM}[CMIM]{Metal-Insulator-Metal Capacitor}
\acrodef{CML}[CML]{Current Mode Logic}
\acrodef{CMOL}[CMOL]{Hybrid CMOS nanoelectronic circuits}
\acrodef{CMOS}[CMOS]{Complementary Metal-Oxide-Semiconductor}
\acrodef{CNN}[CNN]{Convolutional Neural Network}
\acrodef{CNS}[CNS]{central Nervous System}
\acrodef{COTS}[COTS]{Commercial Off-The-Shelf}
\acrodef{CPG}[CPG]{Central Pattern Generator}
\acrodef{CPLD}[CPLD]{Complex Programmable Logic Device}
\acrodef{CPU}[CPU]{Central Processing Unit}
\acrodef{CSM}[CSM]{Cortical State Machine}
\acrodef{CSP}[CSP]{Constraint Satisfaction Problem}
\acrodef{CTXCTL}[CTXCTL]{CortexControl}
\acrodef{CV}[CV]{Coefficient of Variation}
\acrodef{DAC}[DAC]{Digital to Analog Converter}
\acrodef{DAS}[DAS]{Dynamic Auditory Sensor}
\acrodef{DAVIS}[DAVIS]{Dynamic and Active Pixel Vision Sensor}
\acrodef{DBN}[DBN]{Deep Belief Network}
\acrodef{DBS}[DBS]{Deep Brain Stimulation}
\acrodef{DFA}[DFA]{Deterministic Finite Automaton}
\acrodef{DIBL}[DIBL]{Drain-Induced Barrier-Lowering}
\acrodef{DI}[DI]{Delay Insensitive}
\acrodef{divmod3}[DIVMOD3]{Divisibility of a number by three}
\acrodef{DMA}[DMA]{Direct Memory Access}
\acrodef{DNF}[DNF]{Dynamic Neural Field}
\acrodef{DNN}[DNN]{Deep Neural Network}
\acrodef{DOF}[DOF]{Degrees of Freedom}
\acrodef{DPE}[DPE]{Dynamic Parameter Estimation}
\acrodef{DPI}[DPI]{Differential Pair Integrator}
\acrodef{DRAM}[DRAM]{Dynamic Random Access Memory}
\acrodef{DR}[DR]{Dual Rail}
\acrodef{DRRZ}[DR-RZ]{Dual-Rail Return-to-Zero}
\acrodef{DSP}[DSP]{Digital Signal Processor}
\acrodef{DVS}[DVS]{Dynamic Vision Sensor}
\acrodef{DYNAP}[DYNAP]{Dynamic Neuromorphic Asynchronous Processor}
\acrodef{EBL}[EBL]{Electron Beam Lithography}
\acrodef{ECG}[ECG]{Electrocardiography}
\acrodef{ECoG}[ECoG]{Electrocorticography}
\acrodef{EDVAC}[EDVAC]{Electronic Discrete Variable Automatic Computer}
\acrodef{EEG}[EEG]{Electroencephalography}
\acrodef{EIN}[EIN]{Excitatory-Inhibitory Network}
\acrodef{EM}[EM]{Expectation Maximization}
\acrodef{EMG}[EMG]{Electromyography}
\acrodef{EOG}[EOG]{Electrooculogram}
\acrodef{EPSC}[EPSC]{Excitatory Post-Synaptic Current}
\acrodef{EPSP}[EPSP]{Excitatory Post-Synaptic Potential}
\acrodef{EZ}[EZ]{Epileptogenic Zone}
\acrodef{FDSOI}[FDSOI]{Fully-Depleted Silicon on Insulator}
\acrodef{FET}[FET]{Field-Effect Transistor}
\acrodef{FFT}[FFT]{Fast Fourier Transform}
\acrodef{FI}[F-I]{Frequency--Current}
\acrodef{FMA}[FMA]{Floating Microelectrode Array} 
\acrodef{FNN}[FNN]{Feed-forward Neural Network}
\acrodef{FPGA}[FPGA]{Field Programmable Gate Array}
\acrodef{FR}[FR]{Fast Ripple}
\acrodef{FSA}[FSA]{Finite State Automaton}
\acrodef{FSM}[FSM]{Finite State Machine}
\acrodef{GABA}[GABA]{$\gamma$-Aminobutanoic Acid}
\acrodef{GIDL}[GIDL]{Gate-Induced Drain Leakage}
\acrodef{GOPS}[GOPS]{Giga-Operations per Second}
\acrodef{GPIO}[GPIO]{General Purpose I/O}
\acrodef{GPU}[GPU]{Graphical Processing Unit}
\acrodef{GT}[GT]{Ground Truth}
\acrodef{GUI}[GUI]{Graphical User Interface}
\acrodef{HAL}[HAL]{Hardware Abstraction Layer}
\acrodef{HFO}[HFO]{High Frequency Oscillation}
\acrodef{HH}[H\&H]{Hodgkin \& Huxley}
\acrodef{HMM}[HMM]{Hidden Markov Model}
\acrodef{HR}[HR]{Human Readable}
\acrodef{HRS}[HRS]{High-Resistive State}
\acrodef{HSE}[HSE]{Handshaking Expansion}
\acrodef{HW}[HW]{Hardware}
\acrodef{hWTA}[hWTA]{Hard Winner-Take-All}
\acrodef{IC}[IC]{Integrated Circuit}
\acrodef{ICT}[ICT]{Information and Communication Technology}
\acrodef{iEEG}[iEEG]{Intracranial Electroencephalography}
\acrodef{IF2DWTA}[IF2DWTA]{Integrate \& Fire 2-Dimensional WTA}
\acrodef{IF}[I\&F]{Integrate-and-Fire}
\acrodef{IFSLWTA}[IFSLWTA]{Integrate \& Fire Stop Learning WTA}
\acrodef{IMU}[IMU]{Inertial Measurement Unit}
\acrodef{INCF}[INCF]{International Neuroinformatics Coordinating Facility}
\acrodef{INI}[INI]{Institute of Neuroinformatics}
\acrodef{IO}[I/O]{Input/Output}
\acrodef{IoT}[IoT]{Internet of Things}
\acrodef{IP}[IP]{Intellectual Property}
\acrodef{IPSC}[IPSC]{Inhibitory Post-Synaptic Current}
\acrodef{IPSP}[IPSP]{Inhibitory Post-Synaptic Potential}
\acrodef{ISI}[ISI]{Inter-Spike Interval}
\acrodef{JFLAP}[JFLAP]{Java - Formal Languages and Automata Package}
\acrodef{LEDR}[LEDR]{Level-Encoded Dual-Rail}
\acrodef{LFP}[LFP]{Local Field Potential}
\acrodef{LIFE}[LIFE]{Longitudinal Intrafascicular Electrodes}
\acrodef{LIF}[LI\&F]{Leaky Integrate-and-Fire}
\acrodef{LLC}[LLC]{Low Leakage Cell}
\acrodef{LNA}[LNA]{Low-Noise Amplifier}
\acrodef{LPF}[LPF]{Low Pass Filter}
\acrodef{LRS}[LRS]{Low-Resistive State}
\acrodef{LSM}[LSM]{Liquid State Machine}
\acrodef{LTD}[LTD]{Long Term Depression}
\acrodef{LTI}[LTI]{Linear Time-Invariant}
\acrodef{LTP}[LTP]{Long Term Potentiation}
\acrodef{LTU}[LTU]{Linear Threshold Unit}
\acrodef{LUT}[LUT]{Look-Up Table}
\acrodef{LVDS}[LVDS]{Low Voltage Differential Signaling}
\acrodef{MCMC}[MCMC]{Markov-Chain Monte Carlo}
\acrodef{MEA}[MEA]{Multielectrode Arrays}
\acrodef{MEMS}[MEMS]{Micro Electro Mechanical System}
\acrodef{MFR}[MFR]{Mean Firing Rate}
\acrodef{MIM}[MIM]{Metal Insulator Metal}
\acrodef{ML}[ML]{Machine Learning}
\acrodef{MLP}[MLP]{Multilayer Perceptron}
\acrodef{MOSCAP}[MOSCAP]{Metal Oxide Semiconductor Capacitor}
\acrodef{MOSFET}[MOSFET]{Metal Oxide Semiconductor Field-Effect Transistor}
\acrodef{MOS}[MOS]{Metal Oxide Semiconductor}
\acrodef{MRI}[MRI]{Magnetic Resonance Imaging}
\acrodef{NCS}[NCS]{Neuromorphic Cognitive Systems}
\acrodef{NDFSM}[NDFSM]{Non-deterministic Finite State Machine} 
\acrodef{ND}[ND]{Noise-Driven}
\acrodef{NEF}[NEF]{Neural Engineering Framework}
\acrodef{NHML}[NHML]{Neuromorphic Hardware Mark-up Language}
\acrodef{NIL}[NIL]{Nano-Imprint Lithography}
\acrodef{NI}[NI]{Neural Interface}
\acrodef{NMDA}[NMDA]{\textit{N}-Methyl-\textsc{d}-aspartate}
\acrodef{NME}[NE]{Neuromorphic Engineering}
\acrodef{NN}[NN]{Neural Network}
\acrodef{NOC}[NoC]{Network-on-Chip}
\acrodef{NRZ}[NRZ]{Non-Return-to-Zero}
\acrodef{NSM}[NSM]{Neural State Machine}
\acrodef{OR}[OR]{Operating Room}
\acrodef{OTA}[OTA]{Operational Transconductance Amplifier}
\acrodef{PCB}[PCB]{Printed Circuit Board}
\acrodef{PCHB}[PCHB]{Pre-Charge Half-Buffer}
\acrodef{PCM}[PCM]{Phase Change Memory}
\acrodef{PC}[PC]{Personal Computer}
\acrodef{PDK}[PDK]{Process Design Kit}
\acrodef{PE}[PE]{Phase Encoding}
\acrodef{PFA}[PFA]{Probabilistic Finite Automaton}
\acrodef{PFC}[PFC]{Prefrontal Cortex}
\acrodef{PFM}[PFM]{Pulse Frequency Modulation}
\acrodef{PNI}[PNI]{Peripheral Nerve Interface}
\acrodef{PNS}[PNS]{Peripheral Nervous System}
\acrodef{PPG}[PPG]{Photoplethysmography}
\acrodef{PR}[PR]{Production Rule}
\acrodef{PSC}[PSC]{Post-Synaptic Current}
\acrodef{PSP}[PSP]{Post-Synaptic Potential}
\acrodef{PSTH}[PSTH]{Peri-Stimulus Time Histogram}
\acrodef{PV}[PV]{Parvalbumin}
\acrodef{QDI}[QDI]{Quasi Delay Insensitive}
\acrodef{RAM}[RAM]{Random Access Memory}
\acrodef{RA}[RA]{Resected Area}
\acrodef{RDF}[RDF]{Random Dopant Fluctuation}
\acrodef{RELU}[ReLu]{Rectified Linear Unit}
\acrodef{RLS}[RLS]{Recursive Least-Squares}
\acrodef{RMSE}[RMSE]{Root Mean Square-Error}
\acrodef{RMS}[RMS]{Root Mean Square}
\acrodef{RNN}[RNN]{Recurrent Neural Network}
\acrodef{ROLLS}[ROLLS]{Reconfigurable On-Line Learning Spiking}
\acrodef{RRAM}[R-RAM]{Resistive Random Access Memory}
\acrodef{R}[R]{Ripple}
\acrodef{RISC}[RISC]{Reduced Instruction Set Computer}
\acrodef{RSA}[RSA]{Respiratory Sinus Arrhythmia}
\acrodef{SAC}[SAC]{Selective Attention Chip}
\acrodef{SAT}[SAT]{Boolean Satisfiability Problem}
\acrodef{SCI}[SCI]{Spinal Cord Injury}
\acrodef{SCX}[SCX]{Silicon CorteX}
\acrodef{SD}[SD]{Signal-Driven}
\acrodef{SEM}[SEM]{Spike-based Expectation Maximization}
\acrodef{SLAM}[SLAM]{Simultaneous Localization and Mapping}
\acrodef{SNN}[SNN]{Spiking Neural Network}
\acrodef{SNR}[SNR]{Signal to Noise Ratio}
\acrodef{SOC}[SoC]{System-On-Chip}
\acrodef{SOI}[SOI]{Silicon on Insulator}
\acrodef{SOZ}[SOZ]{Seizure Onset Zone}
\acrodef{SP}[SP]{Separation Property}
\acrodef{SPI}[SPI]{Serial Peripheral Interface}
\acrodef{SRAM}[SRAM]{Static Random Access Memory}
\acrodef{SST}[SST]{Somatostatin}
\acrodef{STDP}[STDP]{Spike-Timing Dependent Plasticity}
\acrodef{STD}[STD]{Short-Term Depression}
\acrodef{STP}[STP]{Short-Term Plasticity}
\acrodef{STT-MRAM}[STT-MRAM]{Spin-Transfer Torque Magnetic Random Access Memory}
\acrodef{STT}[STT]{Spin-Transfer Torque}
\acrodef{SVM}[SVM]{Support Vector Machine}
\acrodef{SW}[SW]{Software}
\acrodef{sWTA}[sWTA]{soft Winner-Take-All}
\acrodef{TCAM}[TCAM]{Ternary Content-Addressable Memory}
\acrodef{TFT}[TFT]{Thin Film Transistor}
\acrodef{TIME}[TIME]{Transverse Intrafascicular Multichannel Electrode}
\acrodef{TLE}[TLE]{Temporal Lobe Epilepsy}
\acrodef{UEA}[UEA]{Utah Electrode Array}
\acrodef{USB}[USB]{Universal Serial Bus}
\acrodef{USEA}[USEA]{Utah Slanted Electrode Array}
\acrodef{VHDL}[VHDL]{VHSIC Hardware Description Language}
\acrodef{VHSIC}[VHSIC]{Very High Speed Integrated Circuits}
\acrodef{VIP}[VIP]{Vasoactive Intestinal Peptide}
\acrodef{VLSI}[VLSI]{Very Large Scale Integration}
\acrodef{VNS}[VNS]{Vagal Nerve Stimulation}
\acrodef{VOR}[VOR]{Vestibulo-Ocular Reflex}
\acrodef{VSA}[VSA]{Vector Symbolic Architecture}
\acrodef{WCST}[WCST]{Wisconsin Card Sorting Test}
\acrodef{WTA}[WTA]{Winner-Take-All}
\acrodef{XML}[XML]{eXtensible Mark-up Language}

\author{
Zhe Su, Aron Bencsik, Giacomo Indiveri, Davide Bertozzi

\thanks{
Zhe Su and Giacomo Indiveri are with the Institute of Neuroinformatics, University of Zurich and ETH Zurich. (e-mail: zhesu@ini.ethz.ch).

Aron Bencsik and Davide Bertozzi are with University of Manchester.

This work was supported in part by the Electronic Component Systems for European Leadership (ECSEL) joint undertaking GA No. 876925 (ANDANTE) and in part by the EBRAINS 2.0 project GA No. 101147319 funded by the European Commission and by UKRI Horizon Europe Guarantee Funding.}
}

\maketitle

\begin{abstract}
Multi-core neuromorphic processors are becoming increasingly significant due to their energy-efficient local computing and scalable modular architecture, particularly for event-based processing applications. However, minimizing the cost of inter-core communication, which accounts for the majority of energy usage, remains a challenging issue. Beyond optimizing circuit design at lower abstraction levels, an efficient multicast addressing scheme is crucial. We propose a hierarchical bit string encoding scheme that largely expands the addressing capability of state-of-the-art symbol-based schemes for the same number of routing bits. 
When put at work with a real neuromorphic task, this hierarchical bit string encoding achieves a reduction in area cost by approximately 29\% and decreases energy consumption by about 50\%.
\end{abstract}


\section{Introduction}
\label{sec:intro}

Numerous neuromorphic processors have already demonstrated the benefits of a modular, multi-core architecture \cite{spinnaker_chip, truenorth, loihi, neurogrid, dynapse, braindrop}, highlighting its scalability advantages in light of the physical constraints of individual neural cores. To further enhance the energy efficiency of these processors, it is essential to optimize both power consumption and memory usage when routing spikes between multiple neural cores.

Inspired by biological neurons, where axons branch out to form numerous synaptic connections, the address-event representation (AER) protocol was developed \cite{small-world}. This protocol enables the transmission of neuron addresses to one or more target neural cores, establishing virtual connectivity using configurable lookup tables (LUTs) in both source and target cores. This approach eliminates the need for non-scalable hardwired connections or a full crossbar architecture \cite{memristor}. To implement the AER protocol, established Network-on-Chip (NoC) technologies are utilized.\cite{dynapse,tabula,spinnaker_topology,loihi}.

Efficient target core addressing is key to reducing NoC routing energy and minimizing LUT overhead when managing multicast spike traffic. State-of-the-art approaches to multicast routing in NoCs either result in large packet header overhead (to explicitly address all destination cores) \cite{CMR} or in large routing tables at switching nodes (to store productive outputs for each multicast transaction) \cite{8130788}.
In this paper, we introduce the hierarchical bit string (HBS) encoding scheme for multicast addressing, which outperforms current methods in terms of the number of routing bits required and addressing capability. This scheme is designed to support the development of large-scale, multi-core neuromorphic processors, maintaining low power consumption and memory usage.

Section \ref{sec:related} reviews various multicast addressing schemes utilized in multi-core neuromorphic processors. Section \ref{sec:hbs} presents a theoretical analysis of scalability across these schemes. Section \ref{sec:experiment} describes an experiment involving spike traffic from a real neuromorphic task to validate the effectiveness of the proposed scheme. Finally, conclusions are drawn in Section \ref{sec:conclusion}.

\section{Related work}
\label{sec:related}

\subsection{Multicast in Neuromorphic Processors}
Multi-core neuromorphic processors are distinguished from other ANN accelerators by several key features: (1) fine-grained asynchronous spiking, (2) high temporal sparsity due to inherently dynamic local computation, (3) support for both multicast and unicast traffic, and (4) short payloads (addresses or graded spikes) transmitted over a packet-switched network.

While the NoCs in \cite{spinnaker_chip, neurogrid_noc} support multicast, this functionality is restricted to inter-chip communication. In contrast, \cite{loihi, dynapse} employed on-chip unicast-based multicast by sequentially iterating over the addresses of the destination cores. However, as highlighted in \cite{tabula}, hardware support for parallel tree-based multicast offers lower latency and better energy efficiency than unicast-based multicast, especially when multiple multicast endpoints are involved.

\subsection{Multicast Addressing Encoding}

\begin{figure*}
\centering
\includegraphics[width=4.9in]{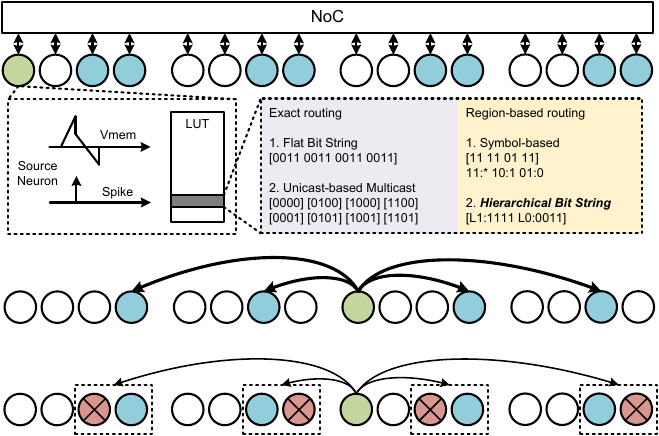}
\caption{Top: A diagram depicting multicast routing in multi-core neuromorphic processors. The green neural core serves as the source, containing one or more source neurons, while the blue neural cores represent the target cores, each holding one or more target neurons. Each "[ ]" represents an entry in the LUT for each source neuron, storing the address of the target core. Middle: Illustrates exact routing, which eliminates the need for filtering illegal data packets but requires a larger packet header (shown as wider arrows). Bottom: Shows region-based routing, where illegal data packets in non-target cores (shown as red cores) must be filtered out.}
\label{fig:comparison}
\end{figure*}

As illustrated in Fig.\ref{fig:comparison} top, whenever a source neuron generates a spike, it triggers a search in the LUT, which contains the multicast address. In parallel tree-based multicast, there is typically one entry per source neuron, whereas, in unicast-based multicast, multiple entries are required for each source neuron during data packet iterations. The multicast address is then used as routing bits in the packet header to determine which output ports should be selected in the crossed NoC switches.

Flat bit string (FBS) encoding is a commonly used format for multicast in exact routing \cite{CMR}. In this scheme, each bit in the address field corresponds to a network node, where a bit is set to 1 if the node is a destination, and 0 otherwise. While this encoding provides flexible targeting for any region (i.e., any combination of neural cores), it results in a header size that scales as $\mathcal{O}(N)$. The scaling is even more significant for unicast-based multicast, where the LUT size grows as $\mathcal{O}(N\log_2{N})$. 

This scaling overhead makes this scheme impractical for neuromorphic computing, especially when the short AER payloads are transmitted through single-word packets where header and payload information is transmitted through parallel wires \cite{tabula}.

A symbol-based encoding scheme, introduced in \cite{jssc2019}, encodes each bit in the address for a set of destination cores using a 2-bit symbol. These symbols can be 0, 1, or *, where the wildcard symbol * indicates that cores with either a 0 or 1 at that bit position are included as multicast destinations. While this method reduces header size, scaling at $\mathcal{O}(\log_2{N})$, it limits the number of possible multicast destination combinations. Essentially, this is a form of region-based routing, which requires the target core to filter out any illegal events \cite{core_interface}.

As illustrated in Fig.\ref{fig:comparison} (middle and bottom), exact routing increases energy consumption per packet transmission due to a larger packet header, while region-based routing requires filtering of illegal events. Greater addressing capability reduces the need for filtering.

This is the challenge we tackle in this paper: we introduce the {\it{HBS multicast addressing encoding scheme, which maintains the same routing bit scaling as symbol-based encoding while largely improving the addressing capabilities.}}

\section{Hierarchical bit string}
\label{sec:hbs}

\subsection{Identifying the optimal hierarchical structure}

\begin{figure}
    \centering
  \subfloat[\label{1a}]{%
        \includegraphics[width=0.5\linewidth]{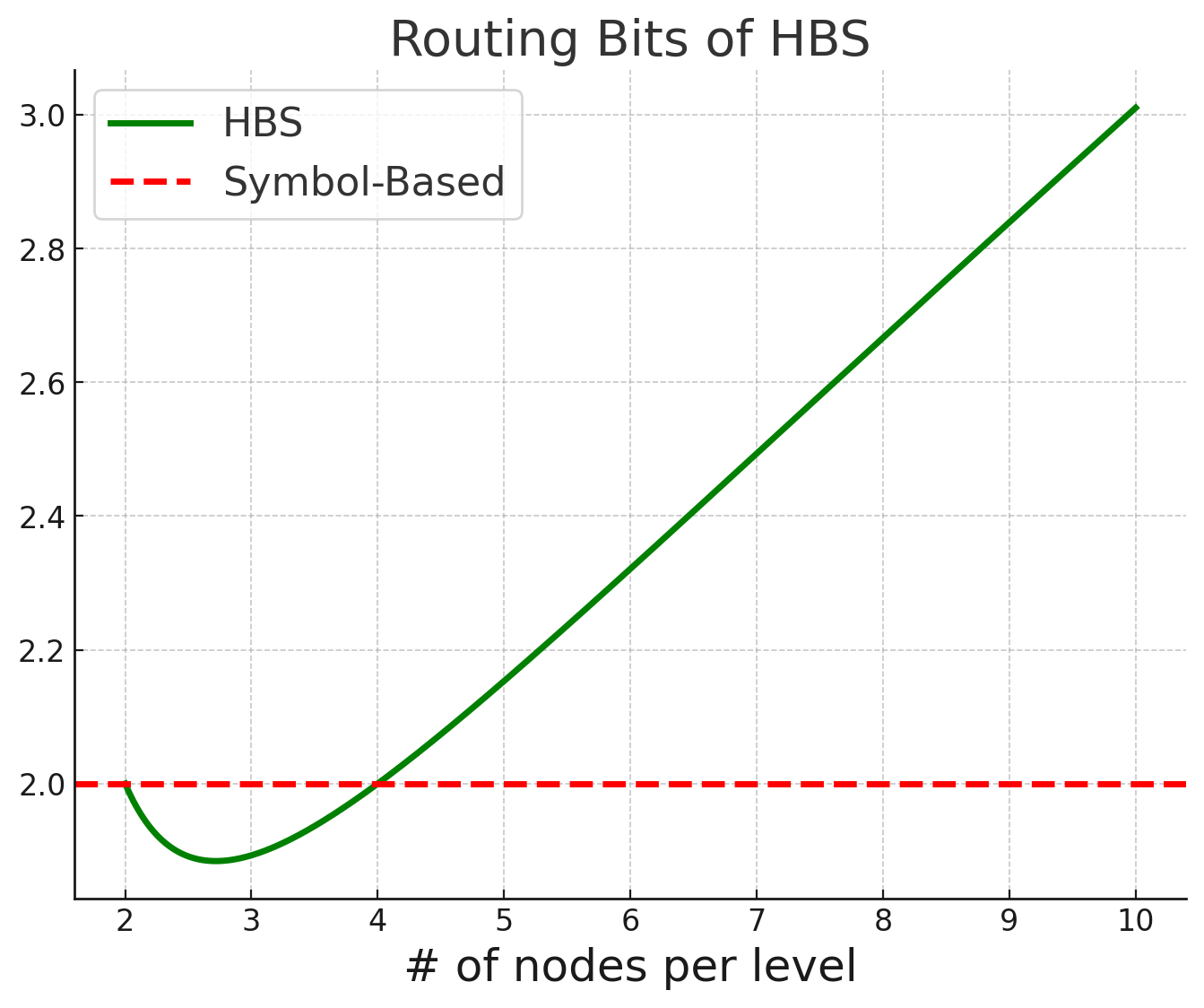}}
    \hfill
  \subfloat[\label{1b}]{%
        \includegraphics[width=0.5\linewidth]{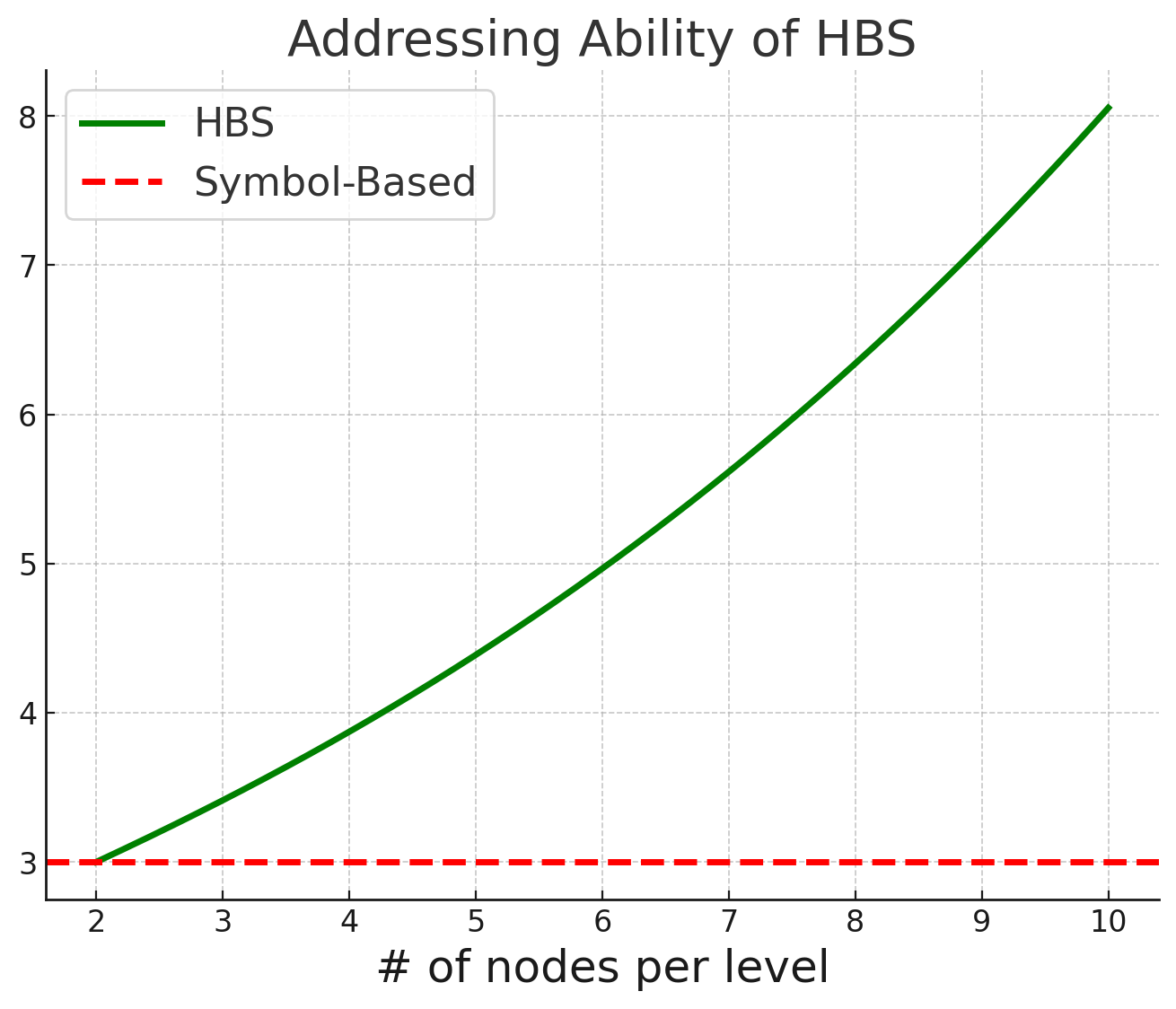}}
  \caption{Determining the number of nodes at each hierarchical level.}
  \label{fig:optimal}
\end{figure}

Instead of using an \( N \)-bit mask to specify each target node in a FBS, the HBS method uses a  \( k \)-bit mask to define a sub-group (or sub-region) at each hierarchical level. In this paper, we assume that the hierarchical levels are evenly distributed, which simplifies designing the routing algorithm and aligns with practical real-world scenarios.
These sub-groups form various multicast trees, and the number of different combinations of these multicast trees is used to evaluate the addressing capability in this study.

Since the second parameter \( k \) is involved, determining its optimal value is necessary before moving into the scalability comparison. As shown in Table. \ref{tab:equation}, there are $\log_k N$ hierarchical levels, with \( k \) bits used at each level. After transforming the equation, the routing bits become proportional to $\frac{k}{\log_2 k}$. This scaling factor is plotted in Fig. \ref{1a}, showing that the optimal integer value of \( k \) is 3. If \( k \) is set to 2 or 4, the scaling factor for routing bits slightly increases to 2, which, interestingly, is similar to symbol-based encoding, where two bits are used to encode three different symbols. However, since 2 or 4 mask bits per level are more compatible with modern computer architectures, we will only consider \( k \)=2 and \( k \)=4 in the following analysis.

When considering the addressing capability, the total number of combinations is calculated by multiplying the number of combinations at each hierarchical level $2^k - 1$ across $\log_k N$ levels. Following a similar approach as described above, we derive the scaling factor $(2^k - 1)^{1/\log_2 k}$ after transformation. As illustrated in Fig. \ref{1b}, the addressing capability increases with the value of \( k \). When \( k \) = 2, the addressing capability corresponds to symbol-based encoding, where each symbol has three possible values. Based on this analysis, the optimal number of nodes per hierarchical level is 4.

\begin{table*}[h!]
\centering
\begin{tabular}{c|c|c|c|c}
\hline
& \textbf{FBS\cite{CMR}} & \textbf{Symbol-based\cite{jssc2019}} & \textbf{HBS} &
\textbf{DYNAPs\cite{dynapse} / Loihi\cite{loihi}} \\ \hline
\textbf{Routing Bits} & $N$ & $2 \log_2 N$ & $\left( \frac{k}{\log_2 k} \right) \log_2 N$* & $N \log_2 N$ \\ \hline
\textbf{Addressing Capability} & $2^N - 1$ & $3^{\log_2 N}$ & $\left((2^k - 1)^{1/\log_2 k} \right)^{\log_2 N}$** & $2^N - 1$ \\ \hline
\textbf{On-chip parallel multicast} & \checkmark & \checkmark & \checkmark & $\times$ \\ \hline
\end{tabular}
\vspace{0.2cm} \\
\small{
* \( k \) denotes the number of nodes at each hierarchical level. $k \log_k N = \left( \frac{k}{\log_2 k} \right) \log_2 N $ \\
** $ \left((2^k - 1)\right)^{\log_k N} = \left((2^k - 1)^{1/\log_2 k} \right)^{\log_2 N}$ }
\caption{Comparison of Multicast Addressing Encoding Schemes}
\label{tab:equation}
\end{table*}

\subsection{A comparison of various multicast addressing encoding schemes}

\begin{figure}
    \centering
  \subfloat[\label{2a}]{%
        \includegraphics[width=0.5\linewidth]{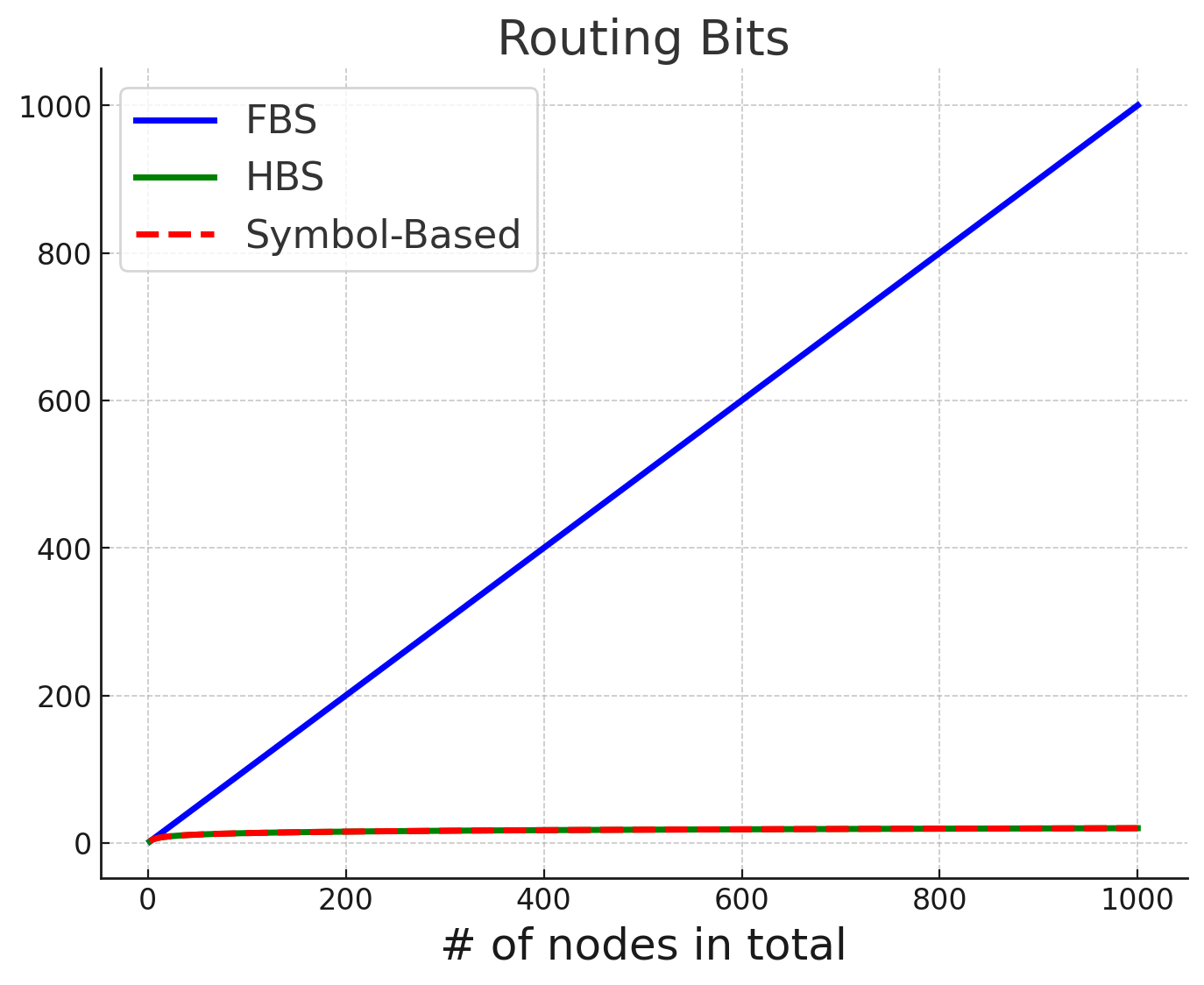}}
    \hfill
  \subfloat[\label{2b}]{%
        \includegraphics[width=0.5\linewidth]{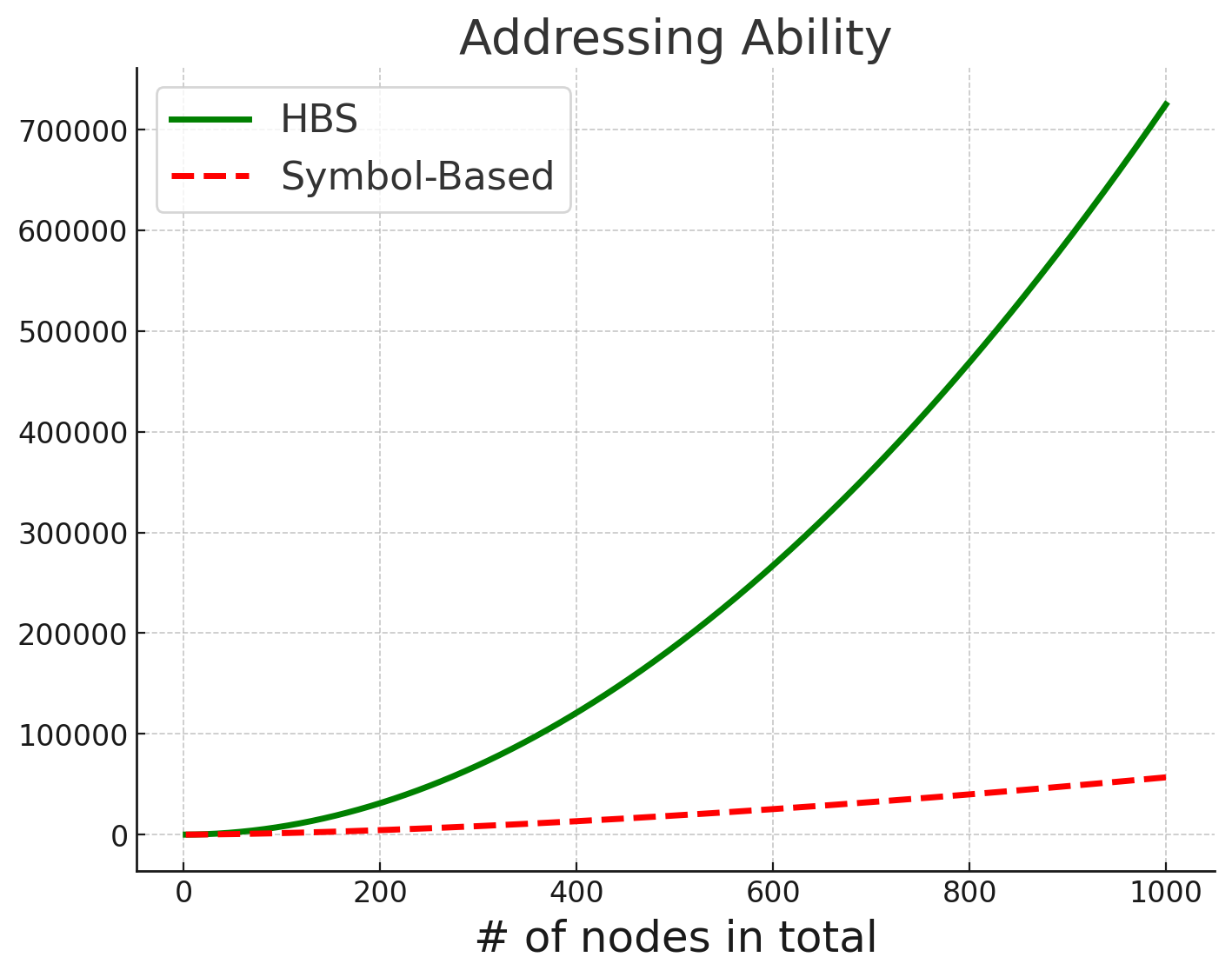}}
  \caption{A comparison of the scalability of various multicast addressing encoding schemes.}
  \label{fig:scalability}
\end{figure}

The scalability is evaluated across various multicast addressing encoding schemes, as illustrated in Table. \ref{tab:equation} and Fig. \ref{fig:scalability}. Table. \ref{tab:equation} also includes unicast-based multicast methods, as used in \cite{dynapse, loihi}, where the scalability of routing bits performs the poorest compared to parallel multicast, resulting in larger LUT sizes and higher routing energy. As depicted in Fig. \ref{2a}, FBS is less scalable compared to HBS and Symbol-based. Moreover, Fig. \ref{2b} demonstrates that HBS significantly improves addressing capability over Symbol-based as the number of computing nodes increases.

\subsection{Hardware mapping}

\begin{figure*}
\centering
\includegraphics[width=4.9in]{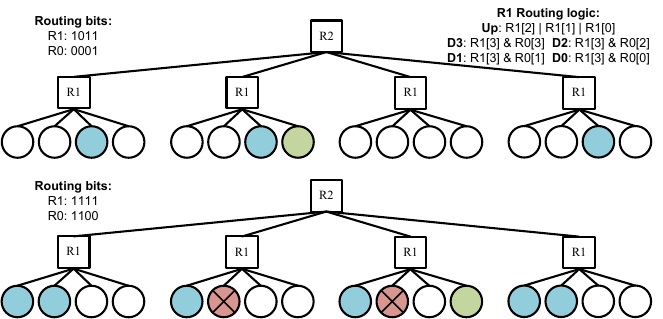}
\caption{HBS mapping in a NoC with a hierarchical tree topology.
Top: The illustration shows an example without illegal multicast. The green core indicates the source core, while the blue cores represent the target cores. The routing logic remains consistent across all R1 data switches, as depicted in the top right corner. \( Up \) signifies selecting the \( Up \) output port for the upward path, and \( D \) indicates choosing the corresponding \( Down \) output port for the downward path.
Bottom: This example demonstrates a scenario with illegal multicast. The red cores highlight those responsible for filtering out illegal spike data packets.}
\label{fig:map}
\end{figure*}

In this section, we explore the mapping of HBS in the NoC. The proposed HBS aligns naturally with the NoC's hierarchical tree topology which has been verified to exhibit exponential scalability in network expansion \cite{hiaer}.



To guarantee that all NoC switches at the same hierarchical level adhere to consistent routing logic in a scalable design, a relative physical address with bit rotation is employed instead of an absolute one. As illustrated in Fig. \ref{fig:map}, the MSB denotes the local branch, while the remaining three bits correspond to sub-nodes based on leftward bit shifting. However, this results in a misalignment with the downward path, where sub-nodes are ordered from MSB to LSB. Bit rotation corrects this alignment by realigning the bits to match the absolute address, without requiring additional hardware.

\section{Experimental results}

\begin{figure}
\centering
\includegraphics[width=0.9\linewidth]{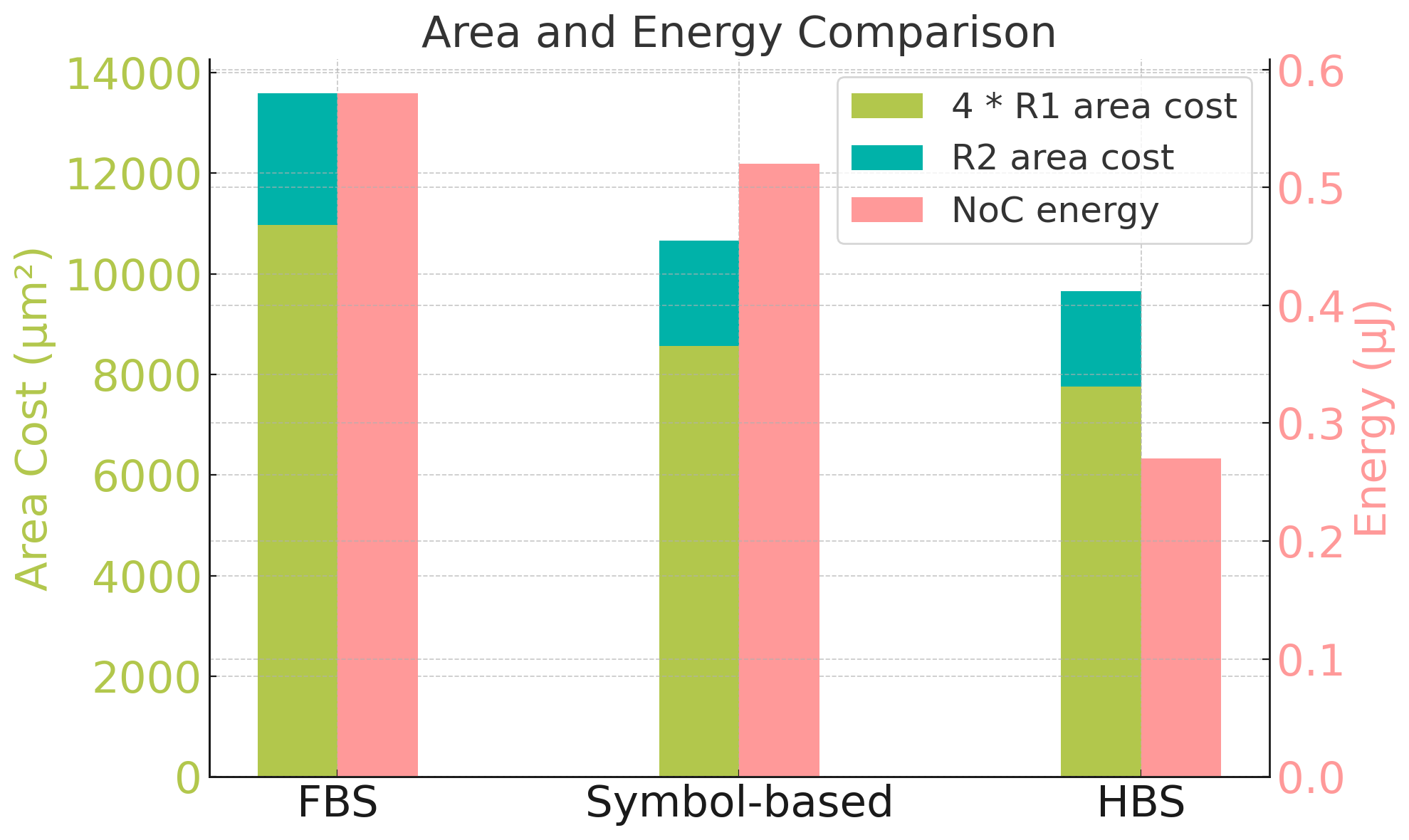}
\caption{The NoC area cost and energy consumption in the real SNN task across various encoding schemes.}
\label{fig:results}
\end{figure}
\label{sec:experiment}

In this section, we conduct an experiment by using the real spike traces from the binary-decision navigation task (NAV), as detailed in \cite{tabula}.

The NoC is built with a hierarchical tree topology, featuring four data switches at the R1 level and one at the R2 level, as shown in Fig. \ref{fig:map}. The switches share the same micro-architecture as in \cite{tabula}, with the routing logic differing for FBS, Symbol-based, and the proposed HBS. The 16 virtual cores, each \( 1 \times 1 \, \text{mm}^2 \) , contain non-routable obstructions but no functional circuits. The NoC is implemented using 22nm technology, with logic synthesis performed in Synopsys DC Shell and place-and-route handled by Cadence Innovus. Power consumption is simulated using QuestaSim and Innovus.

The system implements a Recurrent Spiking Neural Network (RSNN) with three recurrent layers and three fully connected layers of 100 neurons each. Neurons are mapped sequentially to cores within each RSNN layer, switching cores either randomly or when a core reaches its 40-neuron capacity. This mapping is applied randomly 50 times to a 16-core NoC. The spike trace, based on an inference sample of 400 time steps, generates a neuron firing list for each step.

At each time step, target core addresses from source neurons are converted into routing bits based on three multicast encoding schemes. The NoC packet header includes the routing bits and source neuron address tag, which target cores use with a LUT to filter illegal events.

\subsection{Area cost}
The total area cost of the NoC is determined by adding the area costs of the five data switches, while excluding the area of the NoC links with repeaters for simplification. It's noteworthy that all architectures were optimized for the same number of I/O ports (5) and utilized a uniform buffering strategy.
The FBS's bit width is 26 bits, consisting of a 10-bit source tag and 16-bit routing bits. In contrast, both the Symbol-based and HBS approaches use a bit width of 10 bits for the source tag and 8 bits for routing. As illustrated in Fig. \ref{fig:results}, the proposed HBS achieves a 9.4\% reduction in area cost compared to the Symbol-based, thanks to its simpler routing logic, and a 29.0\% reduction compared to FBS, largely due to the significant decrease in bit width, which is a critical factor in assessing the NoC area cost.

\subsection{Energy consumption}
The energy consumption is assessed over the entire NAV sample, which can be split into two components: NoC routing energy and filtering energy from accessing the LUT. The NoC routing energy is illustrated in Fig. \ref{fig:results}. HBS demonstrates a significant energy reduction of 55.7\% (48.1\%) compared to FBS (Symbol-based). This is because, in HBS, illegal transmissions only begin at the R1 switch and follow exact routing at the R2 switch, while Symbol-based starts illegal transmissions at the R2 switch, where the links are much longer than the local ones. FBS incurs higher data routing energy costs due to its larger bitwidth.

In principle, each spike data packet must pass through the filtering LUT upon reaching the target cores, where it is either routed to the appropriate synapse or discarded. This analysis focuses on the energy required to filter out illegal spikes, which varies depending on the encoding schemes used. The number of illegal packets in HBS is less than 30\% of those in the Symbol-based scheme. Energy data from the CAM search in \cite{sota-cam} was referenced, while FBS incurs no filtering energy. In terms of total energy consumption, HBS still achieves a substantial reduction of 52.6\% (49.1\%) compared to FBS (Symbol-based).

\section{Conclusion}
\label{sec:conclusion}
In this paper, we introduced a new addressing method of end nodes for multicast-enabled NoC routing in multi-core neuromorphic processors.
The improved scalability of this scheme is assessed through theoretical analysis and experimental results on a real SNN task. Future work will focus on investigating various network mapping methodologies. Additionally, exploring efficient implementations of axonal or dendritic delays through the NoC will be a key area of interest \cite{delay}.

\printbibliography

\end{document}